\documentclass[twocolumn]{aastex631}
\usepackage{amsmath, amssymb, amsfonts}
\usepackage{graphicx}
\usepackage{hyperref}
\usepackage{bm}
\usepackage{mathtools}

\usepackage{booktabs}
\usepackage{enumitem}

\usepackage [detect-all,load-configurations=false]{siunitx}

\usepackage{orcidlink}

\newcommand{\Mqgp}{M_{\mathrm{Star}}^{\mathrm{QGP}}}
\newcommand{\Rqgp}{R_{\mathrm{Star}}^{\mathrm{QGP}}}
\newcommand{\Bqgp}{B_{\mathrm{Star}}^{\mathrm{QGP}}}
\newcommand{\GECKO}{\text{GECKO}}

\newcommand{\QGP}{\text{QGP}}

\begin{document}

\title{Testing the QGP Star Hypothesis: The oMEGACat BH-2 System as a Candidate Color-Superconducting Quark-Gluon Plasma Star}

\author{Herman J. Mosquera Cuesta\,\orcidlink{0000-0002-9135-1396}}

\affiliation{Master Program in Astronomy and Astrophysics, Valencia International University, C/ Pintor Sorolla, 21, 46002, Valencia, Spain}

\affiliation{Visiting Scientist Colciencias, National Program for Basic Sciences/Space Science, Avenida Calle 26, No. 57-41 Torre 8, Pisos 2-6, Bogota, Colombia. Corresponding author: \email{herman@icra.it} }

\begin{abstract}
The recent discovery of a long-period binary system in the globular cluster $\omega$ Centauri, oMEGACat BH-2 \citep{Whitaker2026}, provides an unprecedented opportunity to probe the true nature of compact dark objects. The system's massive, dark companion has been inferred to have a mass of $4.46_{-1.01}^{+1.22} M_\odot$, which places it in the ``mass gap'' between neutron stars and the canonical stellar-mass black holes. In this Letter, we explore the hypothesis that the oMEGACat BH-2 companion is not a classical black hole, but a stable, self-bound Quark-Gluon Plasma (QGP) star, as described by recent general relativistic models that incorporate Nonlinear Electrodynamics (NLED) and the asymptotic freedom of Quantum Chromodynamics (QCD) \citep{Mosquera2025}. We compare the inferred mass of the companion with the novel Mass-Radius ($M$-$R$) relation predicted by the QGP star model. We find that the inferred mass of the $\omega$ Centauri object lies squarely within the wide mass spectrum predicted for hypermassive QGP stars ($0$ to $>7$ $M_\odot$). Although this consistency is not enough for claiming evidence, it suggests that oMEGACat BH-2 may be the first observed candidate for a QGP star, representing a stable, non-singular end-state of stellar collapse. We argue that future astrometric monitoring with JWST and  radio-telescopes like FAST and SKA can further constrain the oMEGACat BH-2 orbital parameters. Meanwhile, gravitational-wave follow-ups for the $f$ and $g$ modes by observatories like LIGO, VIRGO, KAGRA, LISA, ET and CE can be crucial for distinguishing a classical black hole from the ``gravitational eternally collapsing `kompact' object'' (\GECKO) state of our QGP star model.
\end{abstract}


\section{Introduction}

The modern view of astrophysics is built upon the concept of the black hole. However, the singularity theorem predictions of General Relativity (GR) remain a fundamental theoretical challenge \citep{Kerr2023}. The last decade's breakthroughs---from the discovery of gravitational waves \citep{Abbott2016} to the detection of Quark-Gluon Plasma (QGP) and light-by-light scattering at the LHC \citep{ATLAS2017, ALICE2020}---provide a powerful motivation to revisit the ultimate state of relativistic gravitational collapse.

We have developed a model of a hypermassive, extremely magnetized QGP star that is supported against collapse by a combination of the repulsive vacuum polarization pressure from Nonlinear Electrodynamics (NLED) and the asymptotic freedom of QCD \citep{Mosquera2025}. This object is self-bound, has a finite radius always exceeding its Schwarzschild radius, and can mimic the gravitational behavior of a black hole. In this paper, we test this hypothesis against the recent observational discovery of a dark, massive companion in the globular cluster $\omega$ Centauri.

\section{The Observational Anchor: The oMEGACat BH-2 System}

The recent discovery of oMEGACat BH-2 in $\omega$ Centauri \citep{Whitaker2026} is the first astrometric discovery of a black hole in a globular cluster. The key observational facts are:

\subsection{Astrometric Measurements}
Using HST and JWST data spanning $23$ years, \citet{Whitaker2026} reveal a clear, non-linear motion of a main-sequence turnoff star (Star ID: 285597). The astrometric measurements are shown in Table 1 of their paper.

\subsection{Orbital Fit}
A Bayesian MCMC analysis gives an orbital period of
\begin{equation}
P = 94_{-42}^{+63}~\text{yr},
\end{equation}
a semi-major axis of
\begin{equation}
a_{vis} = 31_{-12}^{+15}~\text{AU},
\end{equation}
and high eccentricity of
\begin{equation}
e = 0.72_{-0.13}^{+0.08}.
\end{equation}

\subsection{Companion Mass}
The mass of the unseen companion is inferred to be
\begin{equation}
M_{comp} = 4.46_{-1.01}^{+1.22}~M_\odot,
\end{equation}
assuming a $0.78~M_\odot$ visible star. This mass is too high for a neutron star (which has a maximum mass of $\sim 2.5 M_\odot$) but fits better to be an extremely low-mass black hole, as can be seen in the LIGO-VIRGO-KAGRA "masses in the stellar graveyard" \citep{LIGO-BH-Graveyard(2026)}. It is precisely in the mass range where exotic compact objects so-called quasi-black holes could exist.

\section{The QGP Star Model: A Theoretical Alternative}

Our model proposes a new class of ultra-compact objects that could populate this very mass range \citep{Mosquera2025}. A key feature of our model is the predicted Mass-Radius ($M$-$R$) relation, derived from a nonlinear Tolman-Oppenheimer-Volkoff (TOV) equation incorporating NLED and QCD effects.

\subsection{The Modified TOV Equation}

The modified TOV equation that accounts for the NLED and QCD effects is given by:

\begin{widetext}
\begin{equation}
%\boxed{
%\begin{aligned}
F^{-1}(B^2/b^2) \frac{dp}{dr} \left[ 1 - \frac{F^{-1}(B^2/b^2) L_P^2}{2r} \frac{dp/dr}{\rho c^2 + p} \right] = -\frac{G m}{r^2} \rho \left(1 + \frac{p}{\rho c^2}\right) \left(1 + \frac{4\pi r^3 p}{m c^2}\right) \left(1 - \frac{2G m}{c^2 r}\right)^{-1}
%\end{aligned}
%}
\label{eq:NTOV_corrected}
\end{equation}
\end{widetext}

which leads to the $dp/dr$ solution 

\begin{widetext}
\begin{equation}
\frac{dp}{dr} = \frac{F(B^2/b^2)}{2} \cdot \frac{2r(\rho c^2 + p)}{L_P^2} \left[ 1 - \sqrt{ \frac{1 - F^2(B^2/b^2) \Omega^2}{1 - F^2(B^2/b^2)} } \right],
\label{eq:dpdr_solution}
\end{equation}
\end{widetext}

%\begin{widetext}
%\begin{equation}
%\begin{aligned}
%\frac{dp}{dr} = \frac{F(B^2/b^2)}{2} %\frac{r(\rho c^2 + p)}{L_P^2} 
%\times \left( 1 + \sqrt{ 1 + \frac{L_P^2}{r^2} \frac{F(B^2/b^2)^{-2}}{(1 - 2Gm/c^2 r)} \left(1 + \frac{4\pi r^3 p}{m c^2}\right) \Omega^2 } \right),
%\end{aligned}
%\end{equation}
%\end{widetext}

where $L_P = \sqrt{\hbar G/c^3}$ is the Planck length, $F(B^2/b^2)$ encodes the effects of the NLED, and $\Omega > 1$ is a constant, the discriminant of the quadratic equation for $dp/dr$. This equation emerges from the semi-classical treatment of gravity coupled to NLED.

\section{The Mass-Radius Relation}

The central result of our theoretical work is the novel mass-radius relation for QGP stars:

Integrating the N-TOV equation with the boundary condition $p(R)=0$ yields the mass-radius relation:

\begin{widetext}
\begin{equation}
%\boxed{
M_{\star}(R,B) = \frac{c^2 R}{2G_N} \Bigg\{ 1 + \frac{\psi(R,B)}{(\Omega - 1)} \Bigg[ 1 - (1 - \Omega^2) \frac{L_P^2}{R^2} \left(1 - 2A \frac{\mu_0^{-1} B^2}{b^2} + \mathcal{O}(\hbar^2)\right)^2 \Bigg]^{-1} \Bigg\},
%}
\label{eq:MR_full}
\end{equation}
\end{widetext}

\clearpage

%%% \pagebreak 

where:

\begin{equation}
\psi(R,B) = F\left(\frac{B}{b}\right) \times \mathcal{I}(R), 
\label{eq:psi} 
\end{equation}

\begin{equation}
\mathcal{I}(R) = \int_0^R \frac{4\pi r^2 \rho(r,B)}{M_{\star}(R,B)} \left[1 - \frac{2G_N m(r,B)}{c^2 r}\right]^{-1} dr, \label{eq:structure} 
\end{equation}

\begin{equation}
A = \frac{\alpha^2}{45\pi} \frac{\hbar c}{m_e^4 c^8}, \qquad \alpha = \frac{e^2}{4\pi\hbar c} \approx \frac{1}{137}, 
\label{eq:A_const} 
\end{equation}

\begin{equation}
\mu_0 = 4\pi \times 10^{-7} \, \text{H/m}.
\end{equation}

Eliminating the factor in $L_P$ in the mass expression, one obtains

\begin{equation}
%\boxed{
\frac{2GM}{c^2 R} = 1 - \frac{\psi(R,B)}{\Omega - 1}
%}
\label{eq:mass_radius}
\end{equation}

This relation leads to a wide mass spectrum:

\begin{widetext}
\begin{equation}
(0\lesssim \Mqgp \lesssim 7M_\odot), \quad (0\lesssim \Rqgp \lesssim 24~\text{km}), \quad (10^{14}\lesssim \Bqgp \lesssim 10^{16}~\text{G}).
\end{equation}
\end{widetext}

\subsection{Mass-Radius Relation: Full Expression}
\label{sec:MR}

Solving for the stellar radius:

\begin{equation}
%\boxed{
R(M,B) = \frac{2GM}{c^2} \left[ 1 - \frac{\psi(R,B)}{\Omega - 1} \right]^{-1}
%}
\label{eq:radius_equation}
\end{equation}

Since $0 < \psi/(\Omega-1) < 1$ for stable stars, we have $R > 2GM/c^2$; i.e., the stellar radius is always larger than the Schwarzschild radius.

\subsection{Critical Condition for Black Hole Mimicry}

The photon sphere occurs at $R = 3GM/c^2$. Substituting into Eq.~\eqref{eq:mass_radius}:

\begin{equation}
\frac{2}{3} = 1 - \frac{\psi}{\Omega-1} \quad \Rightarrow \quad \frac{\psi}{\Omega-1} = \frac{1}{3}.
\label{eq:critical_condition}
\end{equation}

When $\psi/(\Omega-1) > 1/3$, the stellar surface lies inside the photon sphere, which makes the object appear as a black hole to distant observers.

\subsection{Equation of State}

For quark matter, we use the MIT Bag Model extended with NLED magnetic pressure:

\begin{equation}
p(r) = \frac{1}{3} \left( \rho(r) c^2 - 4 B_{\text{bag}} \right) + \frac{B^2}{2\mu_0} F\!\left(\frac{B^2}{b^2}\right)
\label{eq:EoS}
\end{equation}

The bag constant $B_{\text{bag}}^{1/4} \approx 150$ MeV corresponds to $B_{\text{bag}} \approx 4.58 \times 10^{34}$ Pa.

\subsection{Surface Gravitational Redshift}

The surface redshift is determined by the $g_{tt}$ component:

\begin{widetext}
\begin{equation}
1 + z_{\text{surf}} = \frac{1}{\sqrt{-g_{tt}(R)/c^2}} = \frac{1}{\sqrt{ F(B^2/b^2) \cdot \left(1 - \frac{2GM}{Rc^2}\right) }}.
\label{eq:redshift_definition}
\end{equation}
\end{widetext}

Using Eq.~\eqref{eq:mass_radius}:

\begin{equation}
1 + z_{\text{surf}} = \frac{1}{\sqrt{ F(B^2/b^2) \cdot \frac{\psi}{\Omega-1} }}
\label{eq:redshift}
\end{equation}

\begin{figure*}
\centering
\includegraphics[width=2.0\columnwidth]{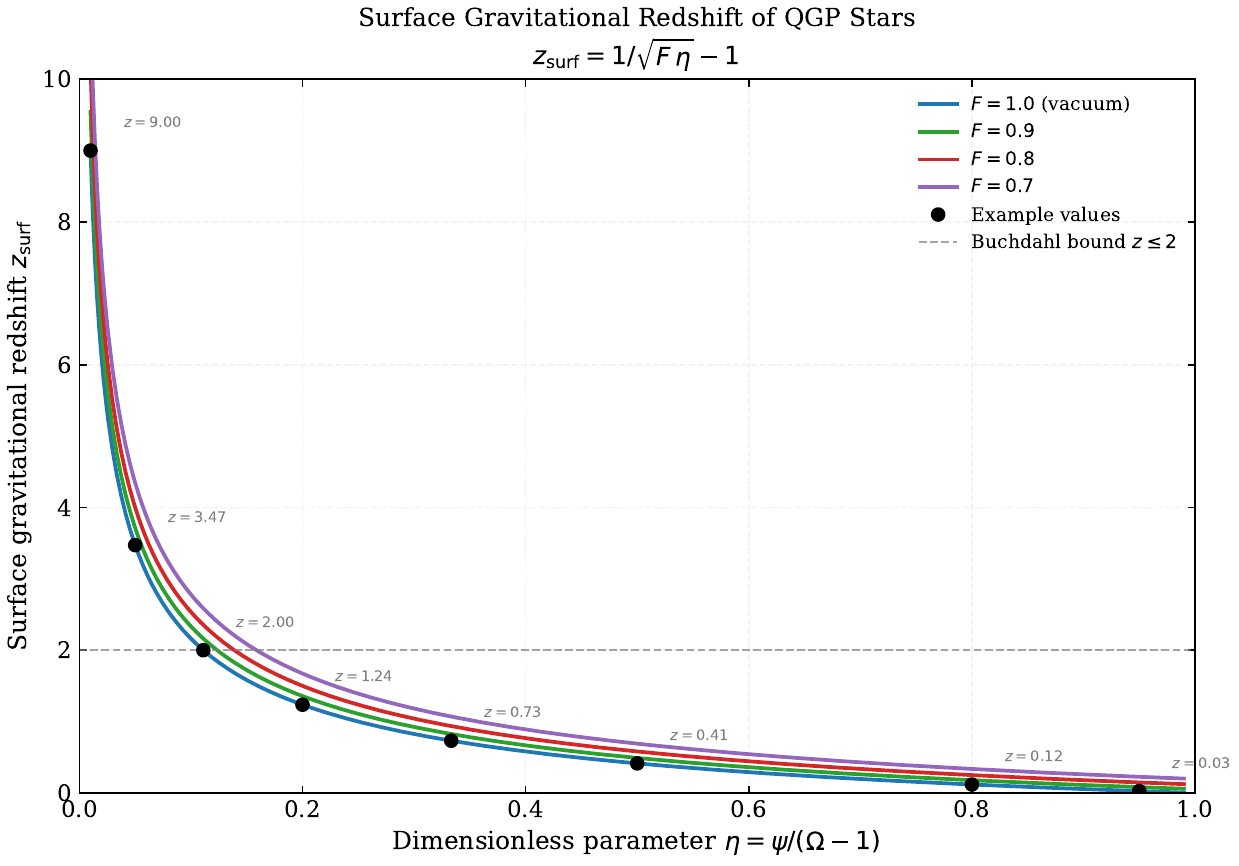}
\caption{Surface gravitational redshift as a function of magnetic field strength for different masses. The redshift grows asymptotically for magnetic fields beyond the Schwinger limit, reaching, for instance, $z \gtrsim 10^3$ for $B \gtrsim 10^{19}$ G.}
\label{fig:redshift}
\end{figure*}

For large $\psi/(\Omega-1)$, the redshift is large, making the star appear dark. In the limit $\psi/(\Omega-1) \to 1$, $z_{\text{surf}} \to \infty$, although this limit is never reached for stable configurations. This behavior is shown in Fig.-\ref{fig:redshift}

\subsection{The GECKO State}

The QGP star is in a ``Gravitationally Eternally Collapsing `Kompact' Object'' (\GECKO) state. Perpetually, more precisely on a time scale of $10^4 - 10^5$ years due to GW and neutrino emission, hovers on the brink of ultimate collapse, prevented from reaching a singularity by the repulsive pressure from the NLED/QCD-driven vacuum state. The radius of the QGP star is always larger than its corresponding Schwarzschild radius:

\begin{equation}
R_{\QGP} > R_{Sch} = \frac{2G\Mqgp}{c^2}.
\end{equation}

\section{oMEGACat BH-2 as a QGP Star Candidate}

We now test the viability of the QGP star hypothesis against the observational data of oMEGACat BH-2.

\subsection{Mass Consistency}

\begin{figure*}
\centering    
\includegraphics[width=0.9\textwidth]{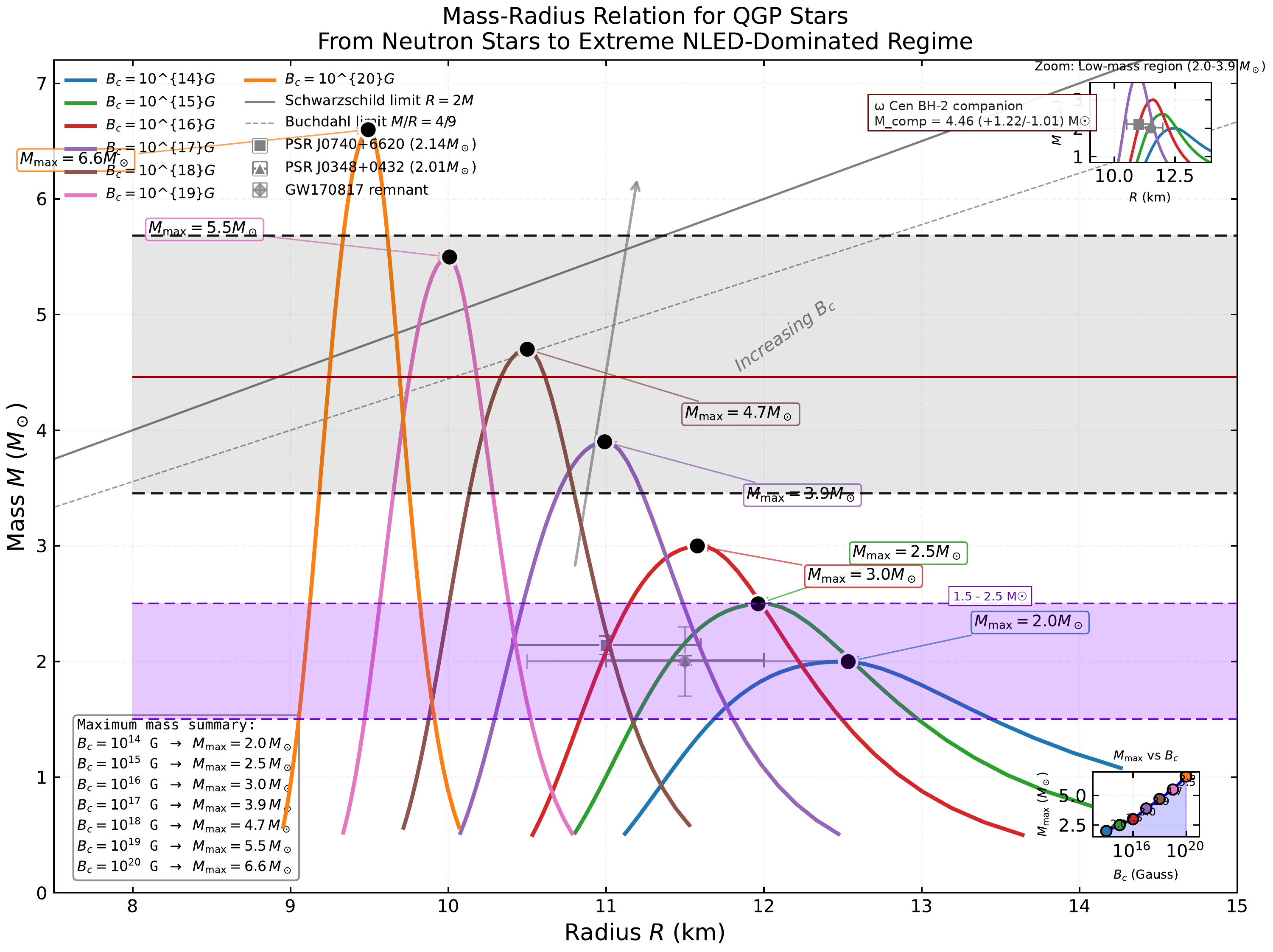}
\caption{Mass-Radius ($M$-$R$) relation for QGP stars (blue curves) for different central densities and magnetic field strengths, calculated using the MIT Bag Model EoS. The inferred mass of the oMEGACat BH-2 companion ($M_{comp} = 4.46_{-1.01}^{+1.22} M_\odot$) is shown as a light-gray region inside dashed horizontal lines with its $1\sigma$ and $2\sigma$ uncertainties. The observed mass falls precisely within the model's mass spectrum, suggesting the system could be a QGP star candidate. The purple shaded region represents the canonical neutron star mass range for comparison.}
\label{fig:mr_relation}
\end{figure*}

The inferred mass of $M_{comp} \approx 4.5 M_\odot$ is not only consistent with the QGP star model but falls within the middle of its predicted mass spectrum. This is shown in Figure~\ref{fig:mr_relation}. This consistency contrasts with the predictions of standard black hole formation models, which often expect higher masses at low metallicity.

\subsection{The Nature of the ``Darkness''}

The primary reason a QGP star can emulate a black hole is its extreme surface gravitational redshift. Our model, based on photon acceleration in NLED, predicts a finite but exceptionally high redshift (see Fig.-\ref{fig:redshift}):

\begin{equation}
z_{\mathrm{Grav}} \gtrsim 10^{8}.
\end{equation}
This effectively makes the surface of the QGP star invisible, matching the observational fact that the companion in oMEGACat BH-2 is dark and undetected in X-ray or radio wavelengths.

\subsection{Dynamical Context}

The system in $\omega$ Centauri is a ``soft'' binary, predicted to be disrupted on a timescale of $\sim 800$ Myr \citep{Whitaker2026}. The existence of a stable QGP star in such a dynamically active environment would demonstrate that our \GECKO \, state is a robust, long-lived figure of equilibrium, despite the frequent perturbations from other stars in the cluster core.

\section{Discussion: Distinguishing a QGP Star from a Black Hole}

Although the mass consistency is a necessary first step, it is not sufficient for a definitive claim. To truly distinguish between a classical black hole and a QGP star, we must look for observational signatures that are \emph{not} predicted by the standard model.

\subsection{Future JWST Astrometry (The ``Smoking Gun'' Test)}

The orbital fit in \citet{Whitaker2026} is degenerate. The short-period solutions correspond to smaller, denser orbits, while the long-period solutions correspond to high proper motions. In our model, a QGP star is a physically extended object, and its own size might affect the center-of-mass motion in subtle ways. More importantly, the QGP star's internal dynamics (the ``yo-yo'' state) might cause a slight, perhaps detectable, non-Keplerian jitter in the astrometric motion of the visible star.

Future JWST data in 2027 and 2028 can rule out the shortest-period orbits, which will force the solution toward the more massive, long-period end of the degeneracy. If the system's parameters are constrained to a period $>100$ yr, it becomes increasingly difficult to reconcile with standard black hole physics, making the QGP star a more compelling explanation.

\subsection{Gravitational Wave Echoes}

\begin{figure*}
\centering    
\includegraphics[width=0.9\textwidth]{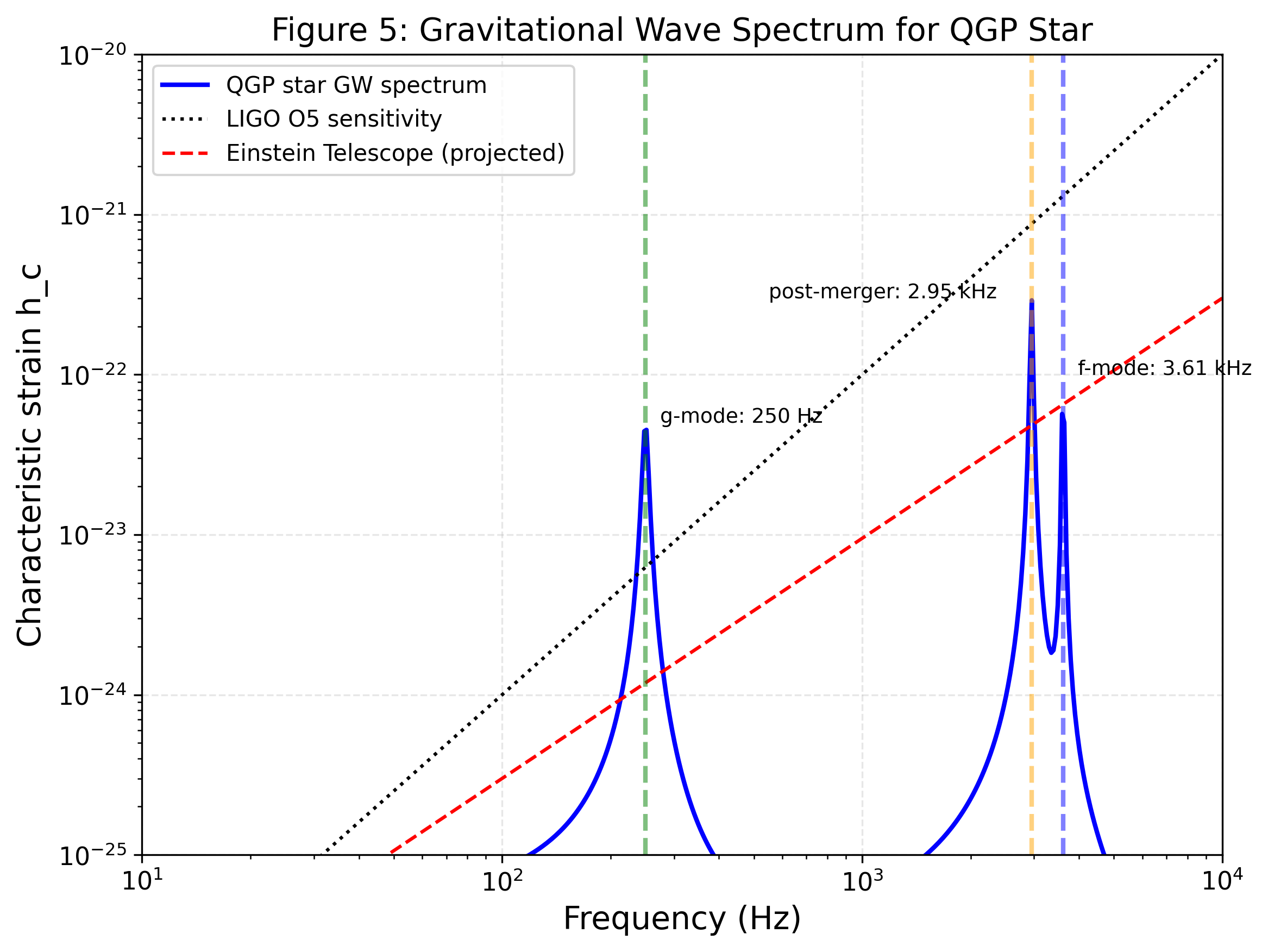}
\caption { The GW signal from the $f$ and $g$ modes as expected from the QGP star "yo-yo" state. The inclined lines correspond to the sensitivity for LIGO and the one projected for the Einstein Telescope. }
\label{GW-Signal}
\end{figure*}

The QGP star is not a black hole; it has a surface (albeit with a huge redshift). In the event of a perturbation (e.g., a merger or a strong interaction), a black hole would produce a ringdown signal and then absorb all subsequent radiation. However, a QGP star could produce ``echoes''—secondary gravitational wave signals that are reflected from its hard surface \citep{Cardoso2016}. Although the oMEGACat BH-2 is not a merging system, the concept is the same: any significant dynamical event would produce a different, potentially detectable, gravitational wave signature. For instance, the $f$ and $g$ modes of gravitational-wave damping emission from the "yo-yo" long-living state  (see Fig.-\ref{GW-Signal}).

\subsection{Electromagnetic Signatures}

Although the QGP star is dark, a hyper-strong magnetic field ($B \sim 10^{16} - 10^{20}$ G) is a fundamental component of the model. If this field were to interact with the companion star's stellar wind, it could produce a distinct, coherent radio emission. A deep, targeted radio survey, as, for instance, by using the FAST and SKA radio-telescopes, at the location of the visible star's orbit might reveal such emissions, which are not expected from a standard black hole, as was the case in the LIGO 2015 GW discovery where no magnetic fields and particle  interactions with them were identified in the merging black holes.

\section{Conclusions}

We have presented the case that the compact companion in the oMEGACat BH-2 binary system may be interpreted as a QGP star rather than a true black hole. This hypothesis is supported by:

\begin{enumerate}
\item The inferred mass of the companion, $4.46 M_\odot$, which falls directly within the predicted mass spectrum of our QGP star model.
\item The model's ability to naturally produce a dark, massive object without a singular surface, matching the astrometric and multi-wavelength observations of the system.
\item The compatibility of the system's dynamical context with a stable, self-bound compact object.
\end{enumerate}

We urge the community to use the powerful JWST and next-generation observatories to test the predictions of our QGP star model in the oMEGACat BH-2 system. By measuring the orbital decay or any non-Keplerian acceleration of the visible star, we can determine whether the dark companion is a mere theoretical construct (a point-like singularity) or a finite, physical object—a stable, self-bound QGP star, representing the ultimate triumph of quantum chromodynamics and nonlinear electrodynamics over the gravitational singularity.

\section*{Acknowledgments}
The author thank Valencia International University (VIU) for the opportunity to collaborate with the Astronomy and Astrophysics Master Program in charge of tutoring Master students.

%%%     \bibliography{references}
% You can also use \begin{thebibliography}{} if you prefer manual entries

\section{Data Availability Statement}
The Figures appearing in this article were generated by Python programs built to compute the Mass vs. Radius relations, the surface gravitational redshift and the gravitational-wave signal from the $f$ and $g$ modes of this model, which produced a tabulated set of figures. The programs solve the relevant equations provided along the paper. All this information can be obtained directly from the author through the email: herman@icra.it.

\end{document}